iv
\documentclass[%
 aip,
 amsmath,amssymb,
 reprint,%
]{revtex4-1}

\usepackage{graphicx}
\usepackage{dcolumn}
\usepackage{bm}

\usepackage[utf8]{inputenc}
\usepackage[T1]{fontenc}
\usepackage{mathptmx}
\usepackage{etoolbox}
\usepackage{hyphenat} 
\usepackage{physics} 
\usepackage{xcolor}
\usepackage{hyperref}
\usepackage{soul}
\usepackage[normalem]{ulem}
\usepackage{siunitx}
\graphicspath{{figures/}} 

\makeatletter
\def\@email#1#2{%
 \endgroup
 \patchcmd{\titleblock@produce}
  {\frontmatter@RRAPformat}
  {\frontmatter@RRAPformat{\produce@RRAP{*#1\href{mailto:#2}{#2}}}\frontmatter@RRAPformat}
  {}{}
}%
\makeatother
\begin{document}

\preprint{AIP/123-QED}

\title[Dispersive readout of a high-Q encapsulated micromechanical resonator]{Dispersive readout of a high-Q encapsulated micromechanical resonator}

\author{Nicholas E. Bousse}
\email{nbousse@stanford.edu}
\affiliation{\mbox{Department of Mechanical Engineering, Stanford University, Stanford, CA 94305, USA}}%

\author{Stephen E. Kuenstner}%
\affiliation{Department of Physics, Stanford University, Stanford, CA 94305, USA}%

\author{James M. L. Miller}
\affiliation{Department of Mechanical Science \& Engineering, University of Illinois at Urbana-Champaign, Urbana, IL 61801, USA}
\author{Hyun-Keun Kwon}
\author{Gabrielle D. Vukasin}
\affiliation{\mbox{Department of Mechanical Engineering, Stanford University, Stanford, CA 94305, USA}}%

\author{John D. Teufel}
\affiliation{National Institute of Standards and Technology, Boulder, Colorado 80305, USA}%

\author{Thomas W. Kenny}
\affiliation{Department of Physics, Stanford University, Stanford, CA 94305, USA}%

\date{\today}

\begin{abstract}
Encapsulated bulk mode microresonators in the megahertz range are used in commercial timekeeping and sensing applications but their performance is limited by the current state of the art of readout methods. We demonstrate a readout using dispersive coupling between a high-Q encapsulated bulk mode micromechanical resonator and a lumped element microwave resonator that is implemented with commercially available components and standard printed circuit board fabrication methods and operates at room temperature and pressure. A frequency domain measurement of the microwave readout system yields a displacement resolution of $522 \, \mathrm{fm/\sqrt{Hz}}$, which demonstrates an improvement over the state of the art of displacement measurement in bulk-mode encapsulated microresonators. This approach can be readily implemented in cryogenic measurements, allowing for future work characterizing the thermomechanical noise of encapsulated bulk mode  resonators at cryogenic temperatures.  
\end{abstract}

\maketitle

Micro\hyp{ }and nanoelectromechanical (M/NEM) resonators are widely used as timing references \cite{leeLowJitterTemperature2012}, inertial sensors \cite{Bogue2007,Shin2017}, and mass sensors \cite{Zhang2005,Kumar2011}. Due to their small size, MEM resonators have high resonance frequencies in the kHz to MHz range and low Size, Weight and Power (SWaP) requirements. Resonators using the bulk modes of encapsulated silicon plates have several additional desirable properties, including mechanical quality factors above 1 million\cite{hamelinMonocrystallineSiliconCarbide2019} and  environmental isolation\cite{kwonOvenControlledMEMSOscillator2019a}, making them extremely promising in precision sensing applications. 

However, this encapsulation process imposes limitations on the resonator sensing methods that can be used. Piezoelectric sensing provides high signal-to-noise ratio but is challenging to incorporate with high-temperature epitaxial encapsulation and introduces significant losses to the resonator, degrading bandwidth and increasing phase noise \cite{pillaiPiezoelectricMEMSResonators2021}. Optical sensing methods offer femtometer displacement resolution but require large, high power, external measurement equipment and are incompatible with encapsulated devices that are opaque to visible light \cite{rembeMeasuringMEMSMotion2019,Gokhale2018}. Capacitive detection relies on coupling between the resonator motion and the field in electrodes placed near the resonator. At higher mechanical frequencies, direct readout can suffer from feed-through, reducing sensitivity and increasing the noise of the position detection \cite{leeDirectParameterExtraction2011}. Optomechanical-style readout is an attractive alternative to direct readout, not only offering improvements in readout noise, but also harnessing decades of progress in the field of quantum optomechanics\cite{Braginski1967,Chan2011,Aspelmeyer2014}. 

In optomechanics, a mechanically compliant structure modulates the resonant frequency of a higher-frequency electromagnetic mode, typically either a resonant microwave circuit or an optical cavity. This modulation upconverts the position information from the mechanical resonance frequency to the microwave/optical frequency, which has enabled position detection near\cite{Teufel2009,Schreppler2014} and even beyond\cite{Aasi2013} the quantum limit in several systems. The electromagnetic mode can also modify the dynamics of the mechanical resonator, which can cool it to its quantum mechanical ground state\cite{Teufel2011} and generate entangled states shared between the electromagnetic and mechanical mode\cite{Palomaki2013}. 

While resonator encapsulation is opaque to visible light, vias through the encapsulation layer allow for optomechanical coupling via lower frequency electromagnetic waves in the megahertz to gigahertz regime. This makes encapsulated bulk mode silicon MEMs devices an excellent basis for further cavity optomechanics experiments, offering extremely high, material-property-limited quality factor and frequency products $(Qf)$\cite{rodriguezDirectDetectionAkhiezer2019} and easy integration into microwave circuits. 

\begin{figure}[!ht]
\includegraphics[width=3.25in]{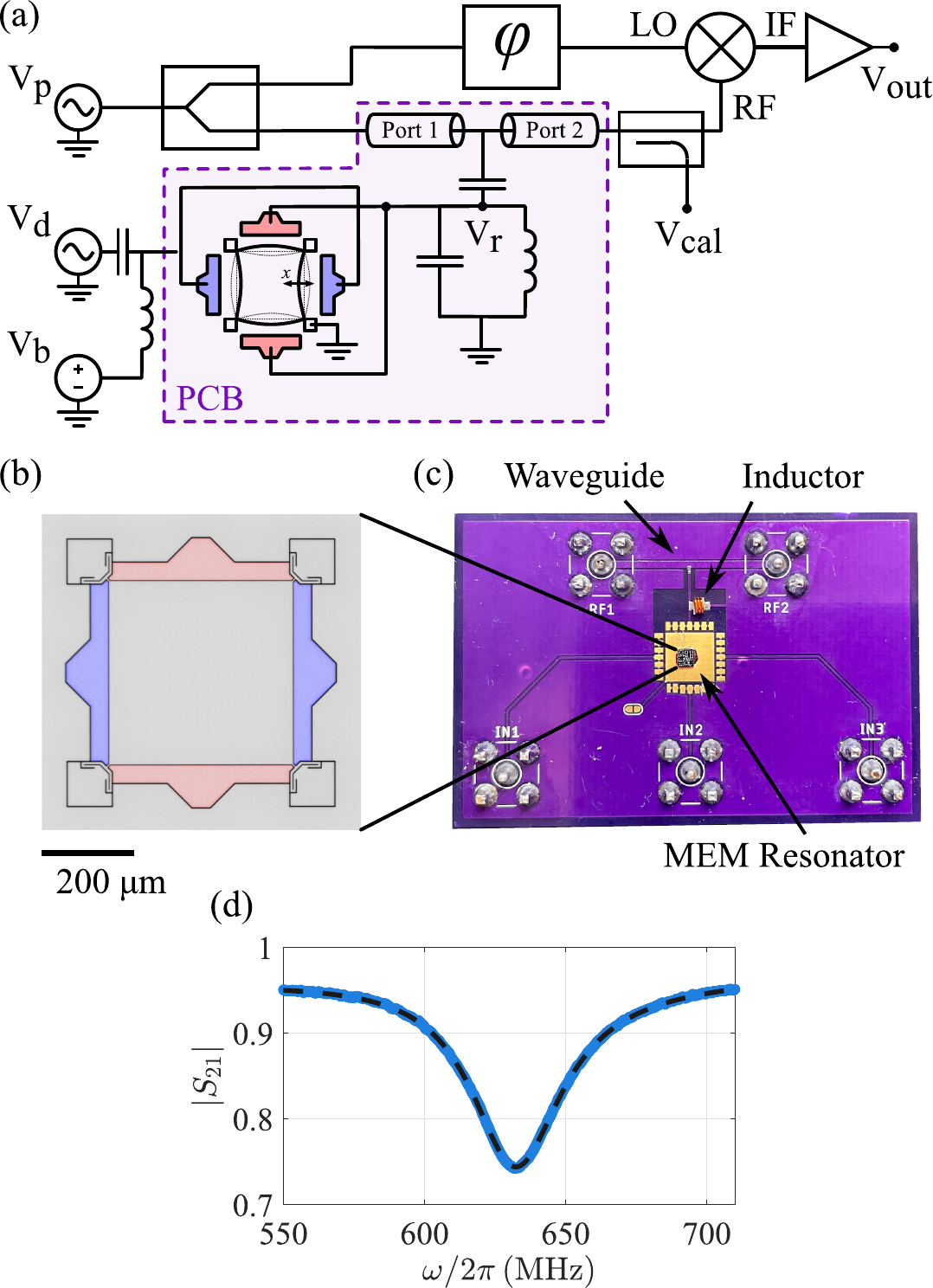}
\caption{\label{fig:setup} (a) A schematic of the coupled resonator system and microwave homodyne readout. Motion of the Lam\'e mode plate resonator, shown in dotted lines, modulates the capacitance between the sense electrode connected to node $\mathrm{V_r}$ and ground. This modulation induces sidebands in the resonator response to a probe voltage, $\mathrm{V_p}$, which are measured using a doubly balanced mixer as a phase detector in a homodyne configuration. The amplified mixer output voltage, $\mathrm{V_{out}}$, is proportional to the mechanical resonator's displacement. A small portion of the resonator signal, $\mathrm{\mathrm{V_{cal}}}$, is measured directly for readout calibration purposes using a directional coupler. (b) A false color micrograph of the Lam\'e mode resonator showing the drive electrodes in blue and sense electrodes in red. (c) A picture of the printed circuit board (PCB) containing the encapsulated mechanical resonator and lumped element microwave resonator. The two SMA ports on the upper portion of the board correspond to ports one and two in the schematic. One of the three lower ports is used to apply the drive signal. The purple outline denotes the PCB extents in the schematic. (d) A plot of the measured microwave resonator transmission amplitude, $\abs{\mathrm{S_{21}}}$, versus frequency. We plot the measured data (blue dots) versus a fit to a Lorentzian (black dotted line) to determine the natural frequency, $\omega_{rf} = 631.6 \,\mathrm{MHz}$, intrinsic quality factor, $Q_I = 38.7$, and  loaded quality factor, $Q_L = 15.0$, of the microwave resonator.}
\end{figure}

In this Letter, we study an encapsulated bulk mode resonator that is dispersively coupled to a lumped-element microwave resonator. In this configuration, motion of the high-Q Lam\'e mode resonator modulates the natural frequency of the microwave resonator, inducing sidebands in its frequency response. This system is analogous to the optomechanical system of a Fabry–Perot cavity with a mechanically compliant mirror, which has been well characterized. Using a microwave homodyne receiver, we use these sidebands to measure the mechanical motion of the bulk mode resonator with state-of-the-art resolution. We derive a model for the dispersive interaction in our system using input-output theory that allows us to predict the sideband amplitude resulting from mechanical motion. We validate this model with experiments on our coupled system, and use fitting to extract the key parameters of our system. 

Figure \ref{fig:setup}a depicts a schematic for the coupled resonator system. The surface-mount air core inductor on the printed circuit board (PCB) seen in Figure \ref{fig:setup}c forms a lumped element microwave resonator with the parasitic capacitance of the coil and the encapsulated MEM device connected in parallel. This microwave resonator is capacitively coupled to an on-board coplanar waveguide forming a transmission mode resonator. Figure \ref{fig:setup}d shows the amplitude response of the microwave resonator. Fitting this response to a Lorentzian allows us to extract the resonator natural frequency, $\omega_{rf}/2\pi = 631.6 \,\mathrm{MHz}$, intrinsic quality factor, $Q_I = 38.7$, and loaded quality factor $Q_L = 15.0$. In this configuration, the resonant frequency of the microwave resonator, $\omega_{rf}$, is a function of the total capacitance between the sense terminal of the mechanical device and ground, $C_t$, and the total inductance, $L_r$, and is given by: 
\begin{equation}
    \omega_{rf} = \frac{1}{\sqrt{L_r \, C_t(x)}}.
    \label{eq:omegar}
\end{equation}
The total capacitance has a static contribution from the parasitic capacitance of the inductor, circuit board, and device interconnects, $C_r$, and a position dependant contribution due to the MEM resonator capacitance, $C_m(x)$. The total capacitance can be written as:
\begin{equation}
    C_t(x) = C_r + C_m(x).
\end{equation}

\begin{figure}[t!]
\includegraphics[width=3.25in]{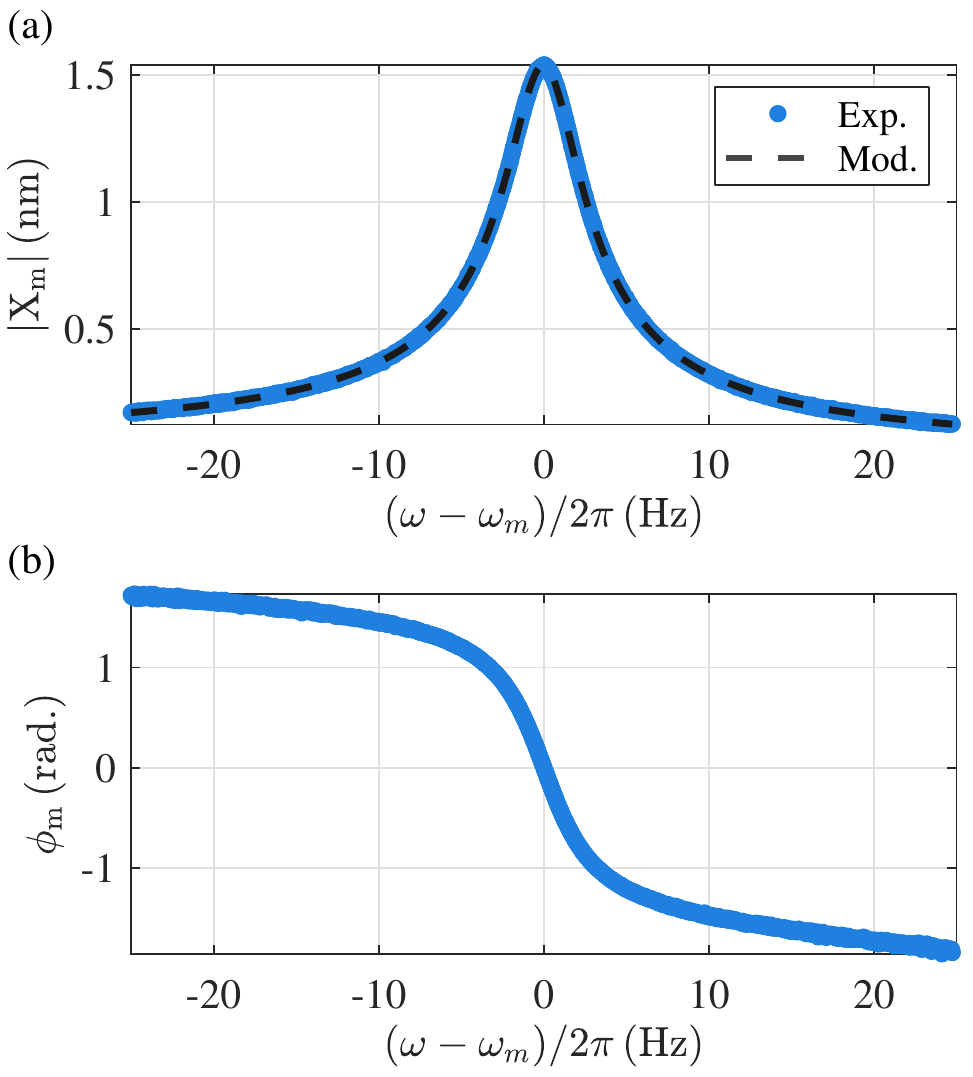}
\caption{\label{fig:OLSweep}The driven amplitude-frequency (a) and phase-frequency (b) response of the Lam\'e mode mechanical resonator measured using the optomechanical readout depicted in Figure \ref{fig:setup} overlaid with a least-squares fit to a Lorentzian response for a drive voltage of 10 mV and a bias voltage of 5 V as a function of offset frequency, $\Delta \omega = \omega - \omega_m$. The good agreement with a Lorentzian model demonstrates the lack of feed-through present in the measurement. This fitting allows us to extract the resonant frequency of the mechanical device as $\omega_m/2\pi = 10.1 \, \mathrm{MHz}$ and the mechanical quality factor as $Q_m = 2.2 \times 10^{\,6}$. The amplitude is calibrated to be in units of meters using the model detailed in the supplemental information. }
\end{figure}

\begin{figure}[t]
\includegraphics[width=3.25in]{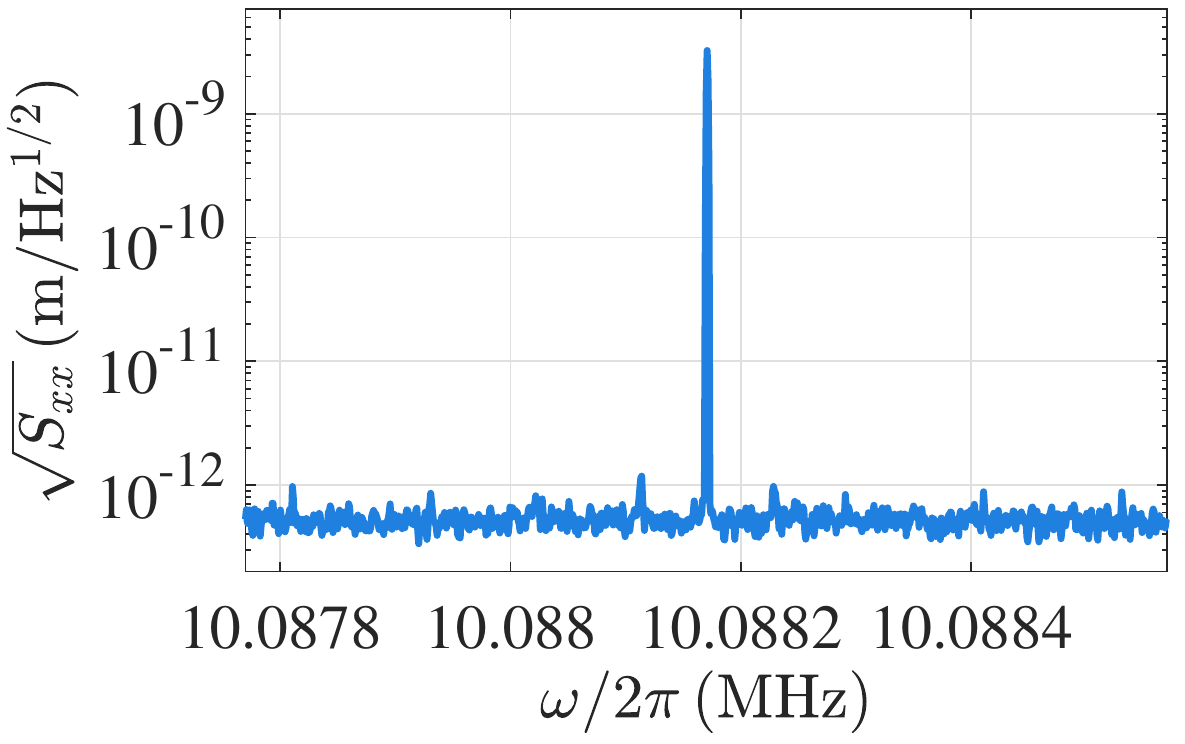}
\caption{\label{fig:spectrum_meters} The amplitude spectral density (ASD) of the driven mechanical resonator as a function of frequency, $\omega/2\pi$, for an applied resonant drive of 30 mV measured using our microwave homodyne detector. The response is calibrated to units of meters  using the calibration detailed in the supplementary material, section IA. Fitting to the white noise floor gives a displacement resolution of $522 \, \mathrm{fm/\sqrt{Hz}}$}. 
\end{figure}

The mechanical resonator is a square, 400 $\mathrm{\mu m}$ wide, 43 $\mathrm{\mu m}$ thick, plate resonator fabricated from single crystal silicon and suspended at the corners with a compliant structure designed to reduce clamping loss via mechanical impedance mismatching \cite{vukasinAnchorDesignAffects2020,riegerEnergyLossesNanomechanical2014}. The resonator is fabricated in a wafer scale encapsulation process that leads to oxide-free, particle-free, low pressure cavities enabling single crystal silicon resonators with high quality factors \cite{yangUnifiedEpiSealProcess2016}. All measurements are performed with the encapsulated mechanical resonator die, lumped-element microwave resonator, and readout electronics inside a temperature-stabilized oven at $\SI{25}\degreeCelsius$ and atmospheric pressure. In this work, we study the resonator's Lam\'e mode, which due to its volume-conserving (isochoric) property exhibits low thermoelastic dissipation \cite{rodriguezDirectDetectionAkhiezer2019}. The lumped mass of this mode is 7.96 $\mathrm{\mu g}$. By applying a periodic drive voltage, $V_d$, offset by a DC bias, $V_b$, through a bias tee, we induce a periodic displacement, $x_m$, in the mechanical resonator given by:
\begin{equation}
    x_m = \abs{X_m(\omega)} \cos\left(\omega t + \phi_m\right),
    \label{eq:resmotion} 
\end{equation}
where $\abs{X_m(\omega)}$ is the magnitude of the resonator's amplitude response due to the applied drive signal and $\phi_m$ is the phase of the resonator. Figure \ref{fig:OLSweep} shows $\abs{X_m(\omega)}$ and $\phi_m$ as the drive signal is swept through the resonant frequency of the mechanical mode. Fitting a Lorentzian to the response yields $\omega_m/2\pi = 10.088 \, \mathrm{MHz}$ and $Q_m = 2.2 \times 10^{\,6}$.
The motion of the mechanical resonator results in time variance of the capacitance between the sense terminal and ground of the mechanical resonator, $C_m$, which is given by:
\begin{equation}
    C_m(x) = \frac{\epsilon_0 \,A}{g-\gamma\, x},
\end{equation}
where $\epsilon_0$ is the permittivity of free space, $A$ is the capacitor area, $g$ is the gap size, and $\gamma$ is the mode shape transduction factor \cite{Kaajakari2009}. This modulation of the capacitance results in modulation of the resonant frequency of the microwave resonator, since $\omega_{rf}$ depends on $C_m$, as can be seen in Eq. \ref{eq:omegar}. The magnitude of the modulation is characterized by the single-photon coupling strength, $g_0$, defined as:
\begin{equation}
    g_0 = \left.\frac{\partial \omega_{rf}}{\partial x}\right\rvert_{x = 0} x_{\mathrm{zpf}},
    \label{eq:g0}
\end{equation}
where $x_{\mathrm{zpf}}$ is the zero point fluctuation of the mechanical mode. The resonant frequency modulation results in a signal at the microwave resonator output at the sum and difference of the frequencies of the microwave resonator and mechanical resonator drive frequencies. These sideband amplitudes from the capacitive modulation are calculated from the current through the microwave resonator capacitance. The current in the microwave resonator, $I_r$,  due to the time variance of the microwave resonator voltage, $V_r$, and the capacitance, $C_t$, is:
\begin{equation}
    I_r =\frac{d}{dt}\left(C_t \,V_r \right).
\end{equation}
\begin{figure}[!t]
\includegraphics[width=3.25in]{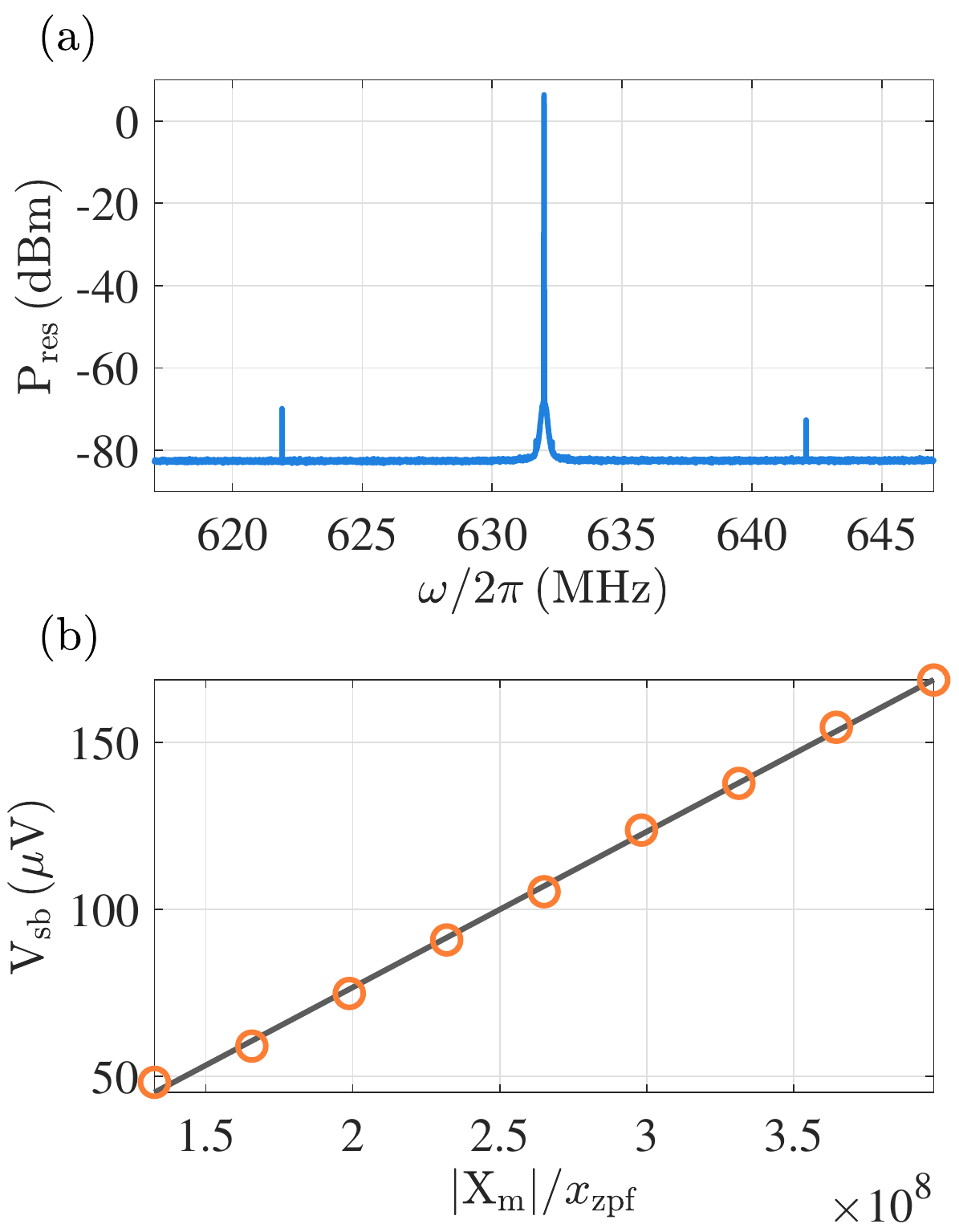}
\caption{\label{fig:calibration} (a) The resonator power plotted versus frequency, $\omega/2\pi$ for a mechanical drive voltage of 30 mV and a bias voltage of 5 V that results in an estimated of displacement amplitude of 4.1 nm. The resonator power was sampled with a resolution bandwidth of 5.1 kHz through a directional coupler and referred back to the microwave resonator output port by applying a known calibration. Motion of the mechanical resonator induces sidebands at $\pm \omega_m$ about the carrier tone proportional to the mechanical resonator amplitude. (b) The total sideband amplitude, $V_{sb}$, plotted versus normalized mechanical resonator amplitude, $\abs{X_m(\omega)}/x_{\mathrm{zpf}}$, for various levels of mechanical resonator drive amplitude. The measured data (open orange circles) are plotted versus a least squares fit to the model in Eq. \eqref{eq:Vsb} (solid grey line), which is used to extract an estimated of the single-photon coupling strength, $g_0/2\pi = 35.3 \,\mathrm{mHz}$.  }
\end{figure}Expanding the total voltage and total capacitance in terms of a sum of the unperturbed value and the slow perturbation induced by the motion of the mechanical resonator:
\begin{equation}
\begin{aligned}
    C_t &= C + \delta C,\\
    V_r &= V_{r0} + \delta V,
\end{aligned}
\end{equation}
where $C$ is the constant capacitance and $\delta C$ is the portion of the capacitance varying at $\omega_m$. $V_{r0}$ is the voltage at $\omega_{rf}$ due to microwave drive power and $\delta V$ is the perturbation at $\omega_m$ due to the capacitive modulation.  Writing the resonator current using these perturbative expansions gives:
\begin{equation}
I_r =\frac{d}{dt}\big(C \,V_{r0} (t)+C \,\delta V (t)+V_{r0} (t)\,\delta C (t)+\delta V (t)\,\delta C (t)\big).
\end{equation}
The first term describes the current in the resonator at $\omega_{rf}$ due to the applied drive power. The second term describes the current induced at $\omega_m$ due to the perturbation $\delta V$ interacting with the constant capacitance. The final term is the product of two small modulation amplitudes and will result in current at $\omega_{rf} \pm 2\omega_m$, which we will ignore in this analysis. The third term creates currents at $\omega_{rf} \pm \omega_m$ due to the product of modulation of the capacitance and the cavity drive. This current is the sideband signal of interest, and results in forward traveling waves at the output port of the resonator, $\tilde{b_2}$. A derivation of this signal can be seen in more detail in the supplementary materials. The magnitude of $\tilde{b_2}$ is given by:
\begin{equation}
\begin{aligned}
\lvert \tilde{b_2}  (\omega_{rf} \pm \omega)\rvert &= \lvert S_{23} \left(\omega_{rf} \pm \omega_m \right)\rvert \\&\frac{ g_0\,\lvert V_r \rvert}{ L_r }\, \frac{\omega_{rf} \pm \omega_m}{\omega_{rf} ^3 }\, \frac{\abs{X_m(\omega)}}{x_{\mathrm{zpf}}} \frac{\sqrt{R_r}}{2},
\label{eq:sidebands} 
\end{aligned}
\end{equation}
where $R_r = \frac{\omega_{rf}\, L_r}{Q_{int}}$ is the equivalent loss resistance, and $\abs{S_{23}}$ is the magnitude of the transmission ratio between the MEM device capacitance and microwave resonator output. We sample the resonator response containing these sidebands directly using a directional coupler. By calibrating the coupler, we  refer the measured voltage at the coupler port, $\mathrm{V_{cal}}$, to resonator output power, $\mathrm{P_{res}}$. Figure \ref{fig:calibration}a shows the spectrum of $\mathrm{P_{res}}$, which displays the carrier tone due to the applied microwave drive at $\omega_{rf}$, and the two sideband signals with amplitude given by Eq. \eqref{eq:sidebands} at $\omega_{rf} \pm \omega_m$. The total sideband voltage, $V_{sb}$, is:
\begin{equation}
V_{sb} = \frac{1}{\sqrt{Z_0}}\left(\lvert \tilde{b_2}  (\omega_{rf} + \omega_m)\rvert + \lvert \tilde{b_2}  (\omega_{rf} - \omega_m)\rvert\right).
\label{eq:Vsb} 
\end{equation}
Fitting to the expression for $V_{sb}$ in Eq. \eqref{eq:Vsb} for various levels of MEM resonator amplitude gives an estimate of the single-photon coupling strength, $g_0/2\pi = 35.3 \,\mathrm{mHz}$. This estimation of the single-photon coupling strength is extremely sensitive to parasitic inductance in the PCB and wirebonds used to connect to the MEM resonator, which is not included in the model. Parasitic inductance can modify the transmission of modulated signals from the MEM resonator to the output, and can modify the  relative circulating power in the capacitor.  The homodyne receiver mixes both of the sideband signals down to low frequency using a doubly balanced mixer as a phase detector and amplifies the resulting voltage. The final output voltage, $V_{out}$, is given by:
\begin{equation}
\begin{aligned}
    \lvert \tilde{V}_{out} (\omega) \rvert &= G_{r} \frac{\sqrt{R_r}}{2\sqrt{Z_0}} \Big[ \lvert S_{23} \left(\omega_{rf} + \omega_m \right) \rvert 
    \\  &+ \lvert S_{23} \rvert\left(\omega_{rf} - \omega_m \right)  \Big]  \frac{ g_0\,\lvert V_r \rvert}{ \omega_{rf}^2 \,L_r}\, \frac{\abs{X_m(\omega)}}{x_{\mathrm{zpf}}},
    \label{eq:vout}
\end{aligned}
\end{equation}
where $G_{r}$ is the net gain of the homodyne detector and amplifier. Figure \ref{fig:spectrum_meters} shows a spectrum of the output voltage given by Eq. \eqref{eq:vout} when a coherent response is induced in the mechanical resonator by applying a bias voltage of $V_b =5\mathrm{\,V}$ and a drive voltage of $V_d =\,30\mathrm{\,mV}$ at the mechanical resonance frequency. This response is calibrated to units of meters using the known applied drive signal and resonator properties, which result in a displacement amplitude of 4.1 nm. For the details of this calibration see the supplementary materials, section IA. The white noise floor gives the displacement resolution of the readout as $522 \, \mathrm{fm/\sqrt{Hz}}$.

Using a Zurich instruments HF2LI lock-in amplifier, we can sweep the frequency of the  MEM resonator drive voltage, $V_d$, and measure the amplitude and phase of the output of the microwave homodyne receiver, $V_{out}$. Calibrating this response to units of meters using the known applied drive signal gives the amplitude response of the mechanical resonator, which can be seen in Figure \ref{fig:OLSweep}. This response, unlike that as measured with direct readout, lacks a direct contribution from the applied drive tone, thereby increasing the stability of clocks and the sensitivity of resonant sensors.

This work demonstrates that encapsulated bulk mode resonators are compatible with an optomechanical-style readout, which offers better noise performance than has been demonstrated with direct readout of encapsulated bulk mode resonators\cite{liDifferentiallyPiezoresistiveTransduction2015}, and comparable noise performance to unencapsulated in-plane optical measurement of RF MEMS \cite{Gokhale2018} and optomechanical measurements of mg-scale silicon resonators \cite{torresHighQualityFactor2013}. Additionally, these resonators are promising for future cavity optomechanics experiments. The $Qf$ product of $2.2 \times 10^{13}$ Hz is comparable with several other leading optomechanics platforms, including SiN nano-resonators \cite{rocheleauPreparationDetectionMechanical2010a,wilsonCavityOptomechanicsStoichiometric2009} and aluminum drumhead resonators \cite{Teufel2011}, and is over the $Qf \gg 6 \times 10^{12}\, Hz$ minimum requirement for room-temperature quantum optomechanics. In these experiments, the mechanical resonance frequency is slightly smaller than the relaxation rate of the microwave resonator, $\gamma_{rf}$ given by $\frac{\omega_{rf}/2\pi}{Q_L} = 34.9\, \mathrm{MHz}$, putting this system in the unresolved sideband regime. However, further optimization of the microwave circuit such as by improving the microwave performance of the vias through the encapsulation layer and including the use of superconducting microwave resonators, can substantially improve the microwave relaxation rate, potentially placing the system into the resolved sideband regime, an important criteria for exploration of quantum phenomena. These improvements would also reduce the system's sensitivity to parasitic inductance and capacitances, and allow for more accurate system characterization. Cryogenic operation also allows for greatly improved rf homodyne detection that will allow for near quantum limited displacement sensitivity \cite{Teufel2009}.

See the supplemental materials for a detailed derivation of the sideband signal amplitude, microwave homodyne receiver behaviour, and the method used for calibrating the MEM resonator response to units of displacement.

This work was supported by the National Science Foundation (NSF) Collaborative Research Program under Grant No. 1662464 and 1662500. Fabrication was performed in nano@Stanford labs, which are supported by the NSF as part of the National Nanotechnology Coordinated Infrastructure under Award No. ECCS-1542152, with support from the Defense Advanced Research Projects Agency’s Precise Robust Inertial Guidance for Munitions (PRIGM) Program, managed by Ron Polcawich and Robert Lutwak.

The data that support the findings of this study are available from the corresponding author upon reasonable request.

This article may be downloaded for personal use only. Any other use requires prior permission of the author and AIP Publishing. This article appeared in N. E. Bousse \textit{et al.} Appl. Phys. Lett. \textbf{121}, 073503 (2022)  and may be found at \url{https://doi.org/10.1063/5.0101402}.  This article is distributed under a Creative Commons Attribution (CC BY) License

\bibliography{Bousse_optomechanics_MEMS_2021_arxiv}

\end{document}


\preprint{AIP/123-QED}

\title{Dispersive readout of a high-Q encapsulated micromechanical resonator: Supplementary material}

\author{Nicholas E. Bousse}
\email{nbousse@stanford.edu}
\affiliation{\mbox{Department of Mechanical Engineering, Stanford University, Stanford, CA 94305, USA}}%

\author{Stephen E. Kuenstner}%
\affiliation{Department of Physics, Stanford University, Stanford, CA 94305, USA}%

\author{James M. L. Miller}
\affiliation{Department of Mechanical Science \& Engineering, University of Illinois at Urbana-Champaign, Urbana, IL 61801, USA}
\author{Hyun-Keun Kwon}
\author{Gabrielle D. Vukasin}
\affiliation{\mbox{Department of Mechanical Engineering, Stanford University, Stanford, CA 94305, USA}}%

\author{John D. Teufel}
\affiliation{National Institute of Standards and Technology, Boulder, Colorado 80305, USA}%

\author{Thomas W. Kenny}
\affiliation{Department of Physics, Stanford University, Stanford, CA 94305, USA}%

\date{\today}

\maketitle
\onecolumngrid

\vspace{-1.2cm}
\section{Sideband Amplitude}
\vspace{-0.2cm}

\begin{figure}[!h]
\centering
\includegraphics[width=6in]{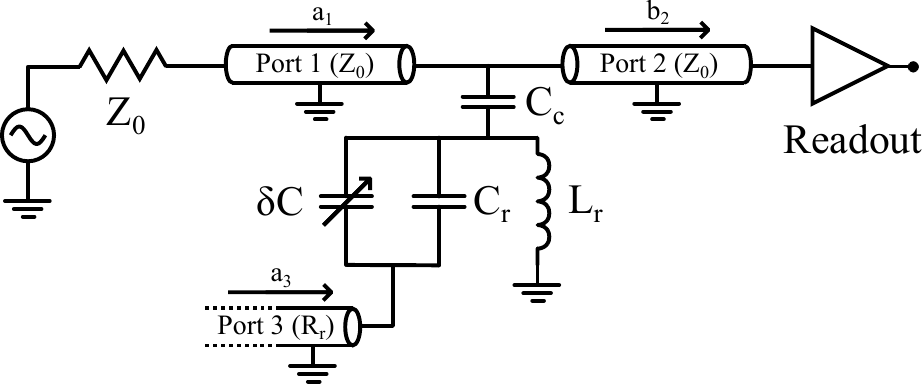}
\setlength{\abovecaptionskip}{0pt}
\setlength{\belowcaptionskip}{0pt}
\caption{Equivalent three port network for the optomechanical system under study. Loss in the microwave resonator is modeled as a unterminated infinitely long transmission line with characteristic impedance $R_r$ connected between the resonator capacitance and ground, which is labled as port three. Port one and two are the resonator input and output port, respectively.  }
\label{3portdiagram}
\end{figure}
The experimental setup seen in Figure 1 of the main text can be can be represented as the three port network seen in Figure \ref{3portdiagram}. Port one is used to drive the resonator, which induces a response that is measured at port two. In this model, the losses in the microwave resonator formed by the lumped capacitance and inductance are modeled as a "loss port", an infinite transmission line with characteristic impedance $R_r$. 

Due to capacitive coupling between the lumped element microwave resonator and encapsulated mechanical resonator, motion of the mechanical resonator induces sidebands in the microwave resonator response. This coupling is due to the dependence of the microwave resonator's natural frequency on the position of the mechanical resonator, and is analogous to the optomechanical system of a Fabry–Perot cavity with a mechanically compliant mirror. Here we will model these sidebands by relating the the mechanical motion to a signal applied to the loss port, which can then be related to the response at the readout port.

Coupling between the mechanical and microwave resonators is achieved by placing the variable capacitance of the mechanical resonator in parallel with the microwave resonator. In this configuration, the microwave resonator's natural frequency is given by:
\begin{equation}
    \omega_{rf} = \frac{1}{\sqrt{L_r \left( C_r + C_m(x)\right)}}
    \label{eq:omegar}
     \end{equation}
where $C_m(x)$ is the position-dependant capacitance of the mechanical resonator, $C_r$ is all other  capacitance, and $L_r$ is the resonator inductance. Motion of the mechanical resonator creates a periodic disturbance in the microwave resonator's natural frequency that can be seen as sidebands in the resonator response seen at port three. The forward propagating wave at port three, $b_3$, which is measured by the microwave homodyne readout, is given by:
\begin{equation}
    b_2 =S_{21} \,a_1 +S_{22} \,a_2 +S_{23} \,a_3,
\end{equation}
where the $S$ is the scattering matrix for the three port network shown in Figure \ref{3portdiagram}. The $a_1$ term is a result of the probe tone applied to the cavity, and lives at $\omega_{rf}$. The $a_2$ term is due to reflection at the readout port, which we will assume to be negligible. The $a_3$ term is the signal due to the fluctuations in the mechanical resonator, which we model as a signal applied to the loss port. The characteristic impedance of the loss port is given by:
\begin{equation}
    R_r = \frac{\omega_{rf}\, L_r}{Q_I},
\end{equation}
where  $Q_I$ is the internal quality factor of the microwave resonator. 

The sideband signal of interest signal is due to the current fluctuations induced by the changing capacitance in the mechanical resonator when it is biased by the circulating voltage in the microwave resonator, $V_r$. This voltage can be found from the circulating power in the microwave resonator, which is given by:
\begin{equation}
    \lvert P_{circ}\rvert =\frac{\lvert P_{\mathrm{inc}}\rvert \,Q_L }{{10}^{{\mathrm{I}\mathrm{L}}/20} \,\pi },
\end{equation}
where $\lvert P_{\mathrm{inc}}\rvert$ is the input power into the resonator, $Q_L$ is the loaded quality factor of the resonator, and $\mathrm{IL}$ is the insertion loss at resonance, in dB. The circulating power can be related to the voltage in the resonator: 
\begin{equation}
    \lvert V_r \rvert =\sqrt{\omega_{rf} L_r \lvert P_{circ} \rvert }.
\end{equation}
As shown in \eqref{eq:omegar}, the total capacitance $C_t$ varies with time due to the motion of mechanical resonator. This can be written in terms of a fixed capacitance and a small perturbation:
\begin{equation}
    C_t =C +\delta C (t),
    \label{eq:C_t}
\end{equation}
where $\delta C$ time varying capacitance due to mechanical motion that creates a small ($\delta C << C$) modulation of the total capacitance at a frequency $\omega_m$ that is much smaller than $\omega_{rf}$. For small resonator amplitudes, this modulation can be approximated as:
\begin{equation}
    \delta C = \left.\frac{\partial C_m}{\partial x}\right\rvert_{x = 0} x_m,
    \label{eq:capmod}
\end{equation}
where $x_m$ is the position of the mechanical resonator. By applying a periodic voltage to the drive electrode of the mechanical resonator, we induce a coherent resonator response which is given by: 
\begin{equation}
    x_m = \abs{X_m(\omega)} \cos\left(\omega_m t + \phi_m\right),
    \label{eq:resmotion} 
\end{equation}
where $\abs{X_m(\omega)}$ is the mechanical resonator's amplitude and $\phi_m$ is its phase. The mechanical resonator's amplitude can be determined based on its known properties and the signal applied to the drive electrode. This derivation is shown in supplemental materials section \ref{disp_model} The coupling between resonator position and capacitance can also be expressed in terms of a coupling between position and frequency shift. The optomechanical coupling rate, $g_0$, is defined as:
\begin{equation}
    g_0 = - \left.\frac{\partial \omega_{rf}}{\partial x}\right\rvert_{x = 0} x_{\mathrm{zpf}},
    \label{eq:g0}
\end{equation}
where $x_{\mathrm{zpf}}$ is the zero point fluctuation of the mechanical mode with effective mass $m_{\mathrm{eff}}$ and natural frequency $\omega_m$ given by:
\begin{equation}
    x_{\mathrm{zpf}}= \sqrt{\frac{\hbar}{2\, m_{\mathrm{eff}}\,\omega_m}}.
\end{equation}
Expanding \eqref{eq:g0} using the expression for the resonator's frequency in \eqref{eq:omegar} gives:
\begin{equation}
    g_0 = \frac{L_r}{2} \omega_{rf}^3 \left.\frac{\partial C_m}{\partial x}\right\rvert_{x = 0} x_{\mathrm{zpf}}.
    \label{eq:g0expanded}
\end{equation}
We can then apply our linearization of $\delta C$ in \eqref{eq:capmod} with our model of resonator motion in \eqref{eq:resmotion}  and expanded optomechanical coupling rate in \eqref{eq:g0expanded} to write an expression for the capacitance modulation in terms of the resonator's amplitude and phase:
\begin{equation}
    \delta C =\frac{2\, g_0}{\omega_{rf} ^3\, L_r} \frac{\abs{X_m(\omega)}}{x_{\mathrm{zpf}}} \, \cos \left( \omega_m t + \phi_m \right).
    \label{eq:capmod_expanded}
\end{equation}
Due to this modulation of the capacitance, the resonator voltage acquires a small perturbation $\delta V$ again at a frequency of $\omega_m$:
\begin{equation}
    V_r =V_{r0} (t)+\delta V (t).
    \label{eq:V_r}
\end{equation}
where $V_{r0}$ is the unperturbed resonator voltage is given by:
\begin{equation}
    V_{r0} (t) = \lvert V_{r} \rvert \cos\left( \omega_{rf} t \right),
\end{equation}
and $\delta V$ is the perturbation voltage given by:
\begin{equation}
    \delta V (t) = \lvert \delta V \rvert \cos\left( \omega_m t + \phi_p\right),
\end{equation}
The fluctuating capacitance and voltage induces a current across the total capacitance given by:
\begin{equation}
    I_r =\frac{d}{dt}\left(C_t \,V_r \right).
\end{equation}
Substituting in the perturbed capacitance and voltage given in \eqref{eq:C_t} and \eqref{eq:V_r} gives: 
\begin{equation}
I_r =\frac{d}{dt}\big(C \,V_{r0} (t)+C \,\delta V (t)+V_{r0} (t)\,\delta C (t)+\delta V (t)\,\delta C (t)\big).
\end{equation}
The last term is the product of two small modulation amplitudes that will result in sidebands at $\omega_{rf} \pm 2\omega_m$ that we will neglect in this analysis. Expanding the first three term gives: 
\begin{equation}
\begin{aligned}
    I_r = -C \,\omega_{rf} \,  \lvert V_r \rvert \sin\left( \omega_{rf} t \right) - C \, \omega_m \,\lvert \delta V \rvert \sin\left( \omega_m t + \phi_p\right) &- \lvert V_r \rvert \cos\left( \omega_{rf} t \right) \frac{2\, g_0}{\omega_{rf}^3 \, L_r} \frac{\abs{X_m(\omega)}}{x_{\mathrm{zpf}}} \,\omega_m\, \sin \left( \omega_m t + \phi_m \right) \\
    &- \omega_{rf} \,\lvert V_r \rvert \sin\left( \omega_{rf} t \right) \frac{2\, g_0}{\omega_{rf}^3 \, L_r} \frac{\abs{X_m(\omega)}}{x_{\mathrm{zpf}}} \, \cos \left( \omega_m t + \phi_m \right).
\end{aligned}
\end{equation}
The first term is the drive tone of the resonator. The second term shows the slow modulation of the resonant frequency due to the capacitance modulation. The final two terms are the sideband signals of interest. Expanding the final two terms gives: 
\begin{equation}
\begin{aligned}
    I_{r,blue} &= -\frac{ g_0\, \abs{X_m(\omega)} \lvert V_r \rvert}{L_r \,x_{\mathrm{zpf}}}  \frac{\omega_{rf} + \omega_m}{\omega_{rf} ^3 } \sin \big( \left( \omega_{rf} + \omega_m \right) t + \phi_m\big),\\
    I_{r,red} &=  -\frac{ g_0\, \abs{X_m(\omega)} \lvert V_r \rvert}{L_r \,x_{\mathrm{zpf}}} \frac{\omega_{rf} - \omega_m}{\omega_{rf} ^3 } \sin \big( \left( \omega_{rf} - \omega_m\right) t -\phi_m\big).
\end{aligned}
\end{equation}
Taking the Fourier transform of this expression, and taking the magnitude gives:
\begin{equation}
\lvert \tilde{I_r} \rvert (\omega_{rf} \pm \omega_m)  = \frac{ g_0\,\lvert V_r \rvert}{ L_r } \, \frac{\abs{X_m(\omega)}}{x_{\mathrm{zpf}}}\,\frac{\omega_m \pm \omega_{rf}}{\omega_{rf} ^3 }.
\end{equation}
This current can be related to the incident power wave at port three, $a_3$:
\begin{equation}
\begin{aligned}
\lvert \tilde{a_3} \rvert (\omega_{rf} + \omega_m) &= \frac{ g_0\,\lvert V_r \rvert}{ L_r }\, \frac{\omega_{rf} + \omega_m}{\omega_{rf} ^3 }\, \frac{\abs{X_m(\omega)}}{x_{\mathrm{zpf}}} \frac{\sqrt{R_r}}{2},\\
\lvert \tilde{a_3} \rvert (\omega_{rf} - \omega_m) &=  \frac{ g_0\,\lvert V_r \rvert}{ L_r }\, \frac{\omega_{rf} - \omega_m}{\omega_{rf} ^3 }\, \frac{\abs{X_m(\omega)}}{x_{\mathrm{zpf}}} \frac{\sqrt{R_r}}{2}.
\end{aligned}
\end{equation}
The incident power wave at port three can be related to the incident power wave into the readout at port two, $b_2$, by the scattering matrix. The resulting power wave at the readout is given by: 
\begin{equation}
\begin{aligned}
\lvert \tilde{b_2} \rvert (\omega_{rf} + \omega_m) &= \lvert S_{23} \rvert\left(\omega_{rf} + \omega_m \right) \frac{ g_0\,\lvert V_r \rvert}{ L_r }\, \frac{\omega_{rf} + \omega_m}{\omega_{rf} ^3 }\, \frac{\abs{X_m(\omega)}}{x_{\mathrm{zpf}}} \frac{\sqrt{R_r}}{2},\\
\lvert \tilde{b_2} \rvert (\omega_{rf} - \omega_m) &= \lvert S_{23} \rvert\left(\omega_{rf} - \omega_m \right) \frac{ g_0\,\lvert V_r \rvert}{ L_r }\, \frac{\omega_{rf} - \omega_m}{\omega_{rf} ^3 }\, \frac{\abs{X_m(\omega)}}{x_{\mathrm{zpf}}} \frac{\sqrt{R_r}}{2}.
\end{aligned}
\end{equation}
where $S_{23}$ is the scattering parameter for transmission from port three to port two, which is derived for this circuit in supplementary material section \ref{S23}.  We can define the total sideband voltage, $V_{sb}$, as the sum of these two sidebands referred to a total voltage:
\begin{equation}
V_{sb} = \frac{1}{\sqrt{Z_0}}\left(\lvert \tilde{b_2}  (\omega_{rf} + \omega_m)\rvert + \lvert \tilde{b_2}  (\omega_{rf} - \omega_m)\rvert\right).
\label{eq:Vsb} 
\end{equation}
Fitting to this expression, shown in Figure 3 of the main text, allows us to extract the coupling rate, $g_0$.

The homodyne receiver shown in Figure 1 of the main text mixes the signal comprising of the carrier tone and these sidebands and mixes with a reference tone. By tuning the phase of the reference tone to achieve zero DC at the IF port, we maximize our phase detector gain. This can be modeled as: 
\begin{equation}
    \lvert \tilde{V}_{out}\rvert (\omega_m) = G_{ro} \frac{1}{\sqrt{Z_0}} \left( \lvert \tilde{b_3} \rvert (\omega_{rf} + \omega_m) + \lvert \tilde{b_3} \rvert (\omega_{rf} - \omega_m)\right),
\end{equation}
where $G_{ro}$ is the readout gain. Expanding this expression gives:
\begin{equation}
    \lvert \tilde{V}_{out}\rvert (\omega_m) = G_{ro} \frac{\sqrt{R_r}}{2\sqrt{Z_0}} \left(\lvert S_{23} \rvert\left(\omega_{rf} + \omega_m \right) + \lvert S_{23} \rvert\left(\omega_{rf} - \omega_m \right) \right)\frac{ g_0\,\lvert V_r \rvert}{ \omega_{rf}^2 \,L_r}\, \frac{\abs{X_m(\omega)}}{x_{\mathrm{zpf}}}.
\end{equation}
This expression provides a model for the output of the microwave readout. A spectrum of $V_{out}$ referred to units of meters using the calibration described in section \ref{disp_model} is shown in Figure 4 of the main text. 

\vspace{-0.2cm}
\subsection{Mechanical Resonator Displacement Modeling}\label{disp_model}
\vspace{-0.2cm}

The mechanical resonator is driven by applying a periodic voltage with a DC offset to an electrode adjacent to the resonator. These applied voltages result in a capacitive potential energy stored in the device electrodes:
\begin{equation}
   U(x,t) = \frac{1}{2} C_m(x) \Delta V(t)^2,
\end{equation}
where $\Delta V$ is the voltage between the resonator and drive electrode that has both a DC component $V_b$ and an AC component $V_d $ applied through a bias tee:
\begin{equation}
    \Delta V(t) = V_b + V_d(t),
\end{equation}
where $V_d(t) = \abs{V_d} \cos(\omega_d t)$ is the drive voltage applied at a frequency of $\omega_d$, which is near the natural frequency of the mechanical resonator. $C_m$ is the capacitance between the resonator and drive electrode that is well modeled by a modified parallel plate capacitor with a capacitance given by:
\begin{equation}
    C_m = \frac{\epsilon_0 \,A}{g-\gamma\, x},
\end{equation}
where $\epsilon_0$ is the vacuum permitivity, $A$ is the area of the plate capacitor plates, $g$ is the gap between the capacitor plates, and $\gamma = .64$ is the mode shape correction factor calculated from the mode shape of the resonator that accounts for the deviation of the capacitance from a pure parallel plate model \cite{Kaajakari2009,Miller2019_JMEMS}. The resulting force on the resonator can be found by taking the positional derivative of the capacitive potential energy evaluated at zero:
\begin{equation}
    f_d =  \left.-\frac{\partial U \left(x, t\right)}{\partial x} \right\rvert_{x = 0} = \left. \frac{\Delta V^2}{2} \frac{\partial C_m}{\partial x} \right\rvert_{x = 0} = -\frac{\epsilon_0 \,A\, \gamma}{2\,g^2} \Delta V^2=-\frac{\epsilon_0 \,A\, \gamma}{2\,g^2} \left(V_b^2+  2 \,V_b\, \abs{V_d} \cos\left(\omega_d t\right) +\abs{v_d}^2 \cos^2\left(\omega_d t\right)\right).
\end{equation}
The term at DC is absorbed into the definition of position, and the term at $2 \omega_d$ is well above the cutoff frequency of the resonator and can be ignored. The relevant drive force near resonance is then given by:
\begin{equation}
F_d =-\frac{\epsilon_0 \,A\, \gamma}{\,g^2}  \,V_b\, \abs{V_d} \cos\left(\omega_d t\right).
\end{equation}
Taking the Fourier transform, we can write this force in the frequency domain:
\begin{equation}
\tilde{F_d}(\omega) =-\frac{\epsilon_0 \,A\, \gamma}{\,g^2}  \,V_b\, \tilde{V_d}(\omega).
\label{eq:driveforce}
\end{equation}
To relate this force to a resonator displacement, we must consider the dynamics of the mechanical system. The equation of motion for the mechanical resonator is given by:
\begin{equation}
    \frac{\partial^2 x}{\partial t^2} + \frac{\omega_m}{Q_m} \frac{\partial x}{\partial t}+ \omega_m^2 x = \frac{F_d}{m_{\mathrm{eff}}},
\end{equation}
where $m_{\mathrm{eff}}$ is the effective modal mass, $\omega_m$ is the natural frequency of the resonator, and $Q_m$ is the quality factor of the resonator. In the frequency domain, the resonator transfer function from force to position is then:
\begin{equation}
H^2(\omega) = \frac{\tilde{x}_d}{\tilde{F_d}} = \frac{ 1 }{m_{\mathrm{eff}}^2 \left[\left(\omega_{m}^2 - \omega^2 \right)^2 + \left( \frac{\omega \omega_{m}}{Q_{m}} \right)^2 \right] }.
\label{eq:tf}
\end{equation} 
\begin{table}[t]
\caption{Summary of mechanical resonator properties and applied signals. }
\label{tab:dev_params}
\begin{tabular}{l|l|l}
\multicolumn{1}{c|}{Parameter} & \multicolumn{1}{c|}{Description}      & \multicolumn{1}{c}{Value}                              \\ \hline
$A$                            & capacitor area                        & $2 \times 400 \mathrm{\mu m} \times 40 \mathrm{\mu m}$ \\
$g$                            & nominal capacitor gap size            & $1.2 \, \mathrm{\mu m}$                                \\
$m_{eff}$                      & effective modal mass                  & $7.96 \,\mathrm{\mu g}$                                \\
$\omega_m$                     & mechanical resonant frequency         & $10.088 \times 2 \pi \, \mathrm{MHz}$                  \\
$Q_m$                          & mechanical quality factor             & $2.2 \times 10^{\,6}$                                  \\
$V_b$                          & DC bias voltage                       & $5 \, \mathrm{V}$                                      \\
$\abs{V_d}$                    & drive voltage magnitude, zero to peak & $5 - 30 \, \mathrm{mV}$                               
\end{tabular}
\end{table}
This transfer function filters the drive force, $F_d$, to yield the driven displacement $x_d$:
\begin{equation}
\tilde{x_d}^2\left(\omega\right) = H^2(\omega) \tilde{F_{d}}^2(\omega).
\end{equation}
Substituting in the drive force in \eqref{eq:driveforce} and the resonator transfer function in \eqref{eq:tf}:
\begin{equation}
    \tilde{x}_d(\omega) = \frac{ \epsilon_0 \,A\, \gamma  \,V_b\, \tilde{V_d}(\omega)}{g^2 \, m_{\mathrm{eff}} \sqrt{\left(\omega_{m}^2 - \omega^2 \right)^2 + \left( \frac{\omega \omega_{m}}{Q_{m}} \right)^2 } }.
\end{equation}
For a $V_d$ given by $V_d(t) = \abs{V_d} \cos(\omega_d t)$, the resulting resonator motion in the time domain is:
\begin{equation}
    x_d(t) = \abs{X_m(\omega_d)} \cos(\omega_d t + \phi_m),
\end{equation}
where $\abs{X_m(\omega_d)}$ is the mechanical resonator amplitude given by:
\begin{equation}
    \abs{X_m(\omega_d)} = \frac{ \epsilon_0 \,A\, \gamma  \,V_b\, \abs{V_d}}{g^2 \, m_{\mathrm{eff}} \sqrt{\left(\omega_{m}^2 - \omega_d^2 \right)^2 + \left( \frac{\omega \omega_{m}}{Q_{m}} \right)^2 } }.
\end{equation}
When the driving force is on resonance ($\omega_d = \omega_m$) this simplifies to:
\begin{equation}
    \abs{X_m(\omega)}_{\omega_d = \omega_m} = \frac{ \epsilon_0 \,A\, \gamma  \,V_b\, \abs{V_d} Q_m}{g^2 \, m_{\mathrm{eff}} \,\omega_m^2 }.
\end{equation}
This expression relates the measured resonator properties ($Q_m$, $\omega_m$), parameters derived from modeling of the mode shape ($m_{\mathrm{eff}}$, $\gamma$), device geometry ($A$, $g$), and applied voltages ($V_b$, $\abs{V_d}$) to the amplitude of resonator motion. While we cannot directly measure the area and gap size of the encapsulated resonator under study, The Bosch-Stanford EpiSeal process used to fabricate this resonator has been extensively characterized, and the realized dimensions are well known. The device parameters needed to calculate resonator amplitude are shown in Table \ref{tab:dev_params}.

\vspace{-0.2cm}
\subsection{Scattering Matrix for Three Port Resonator Model}\label{S23}
\vspace{-0.2cm}

The impedance matrix of the three port resonator model shown in Figure 
\ref{3portdiagram} is given by:
\begin{equation}
\mathbf{Z}(\omega) = 
\begin{bmatrix}
L_r\,\omega\,i-\frac{i}{C_c\,\omega} &L_r\,\omega\,i-\frac{i}{C_c\,\omega}&L_r\,\omega\,i \\
L_r\,\omega\,i-\frac{i}{C_c\,\omega}  & L_r\,\omega\,i-\frac{i}{C_c\,\omega} & L_r\,\omega\,i \\
L_r\,\omega\,i& L_r\,\omega\,i& L_r\,\omega\,i-\frac{i}{C_t\,\omega} 
\end{bmatrix}.
\label{eq:impmatrix}
\end{equation}
 This impedance matrix can be converted to a scattering matrix by applying the following matrix transformation \cite{Ahn2006}:

\begin{equation}
    \mathbf{S} = \mathbf{G_{ref}}^{-1} \left(\mathbf{Z} + \mathbf{Z_{ref}} \right)^{-1} \left( \mathbf{Z} - \mathbf{Z_{ref}}\right) \mathbf{G_{ref}}, \label{eq:scatteringtranform}
\end{equation}
where $\mathbf{Z_{ref}}$ is the diagonal reference impedance matrix given in this case by:
\begin{equation}
    \mathbf{Z_{ref}} =
\begin{bmatrix}
Z_0 & 0 & 0 \\
0 & Z_0 & 0 \\
0 & 0 & R_r 
\end{bmatrix},
\end{equation}
and $\mathbf{G_{ref}}$ is the diagonal matrix given in this case by:
\begin{equation}
    \mathbf{G_{ref}} =
\begin{bmatrix}
\frac{1}{\sqrt{Z_0}} & 0 & 0 \\
0 & \frac{1}{\sqrt{Z_0}} & 0 \\
0 & 0 & \frac{1}{\sqrt{R_r}} 
\end{bmatrix}.
\end{equation}
Applying \eqref{eq:scatteringtranform} to \eqref{eq:impmatrix} gives the scattering matrix for the three port resonator network. The $S_{23}$ term is given by:

\begin{equation}
    S_{23} =
-\frac{2\,C_{c}\,C_{t}\,L_{r}\,\sqrt{R_{r}}\,\sqrt{Z_{0}}\,\omega ^3}{2\,C_{t}\,R_{r}\,\omega +C_{c}\,Z_{0}\,\omega +2\,C_{c}\,L_{r}\,\omega ^2\,\mathrm{i}+2\,C_{t}\,L_{r}\,\omega ^2\,\mathrm{i}-2\,C_{c}\,C_{t}\,L_{r}\,R_{r}\,\omega ^3-C_{c}\,C_{t}\,L_{r}\,Z_{0}\,\omega ^3+C_{c}\,C_{t}\,R_{r}\,Z_{0}\,\omega ^2\,\mathrm{i}-2\mathrm{i}}
\label{eq:S23}
\end{equation}
The parameters for the system seen in Figure 1 of the main text are given in Table 
\ref{tab:elec_params}. Evaluating Eq. \eqref{eq:S23} at the relative sideband frequencies using these parameters gives $S_{23}(\omega_{rf} + \omega_m) = 0.134$ and $S_{23}(\omega_{rf} - \omega_m) = 0.221$

\begin{table}[h]
\caption{Summary of microwave resonator parameters.}
\label{tab:elec_params}
\begin{tabular}{l|l|l}
\multicolumn{1}{c|}{Parameter} & \multicolumn{1}{c|}{Description}     & \multicolumn{1}{c}{Value} \\ \hline
$C_c$                          & Coupling capacitance                 & $1 \, \mathrm{pF}$        \\
$C_t$                          & Total resonator capacitance          & $8.28 \, \mathrm{pF}$     \\
$C_m$                          & Mechanical resonator capacitance     & $0.24 \, \mathrm{pF}$     \\
$L_r$                          & Total resonator inductance           & $8 \, \mathrm{nH}$        \\
$R_r$                          & Equivalent resonator loss resistance & $553 \, \mathrm{m \Omega}$ \\
                               &                                      &                          
\end{tabular}
\end{table}

\bibliography{aux_refs, _zotero}%